\documentclass[useAMS,usenatbib]{mn2e}

\usepackage{graphicx}

\title[Radial migration of the Sun in galactic disk]
  {Radial migration of the Sun in galactic disk}
\author[J.~Kla\v{c}ka et al.]
  {J.~Kla\v{c}ka, M.~Jur\v{c}i, J.~Dur\v{z}o and R. Nagy \\
  Faculty of Mathematics, Physics, and Informatics, Comenius University, Mlynsk\'{a} dolina, 842 48 Bratislava, Slovakia \\
e-mails: klacka@fmph.uniba.sk ; jurci@fmph.uniba.sk ; durzo@fmph.uniba.sk ; roman.nagy@fmph.uniba.sk}

\date{}

\pagerange{\pageref{firstpage}--\pageref{lastpage}} \pubyear{2012}

\def\LaTeX{L\kern-.36em\raise.3ex\hbox{a}\kern-.15em
    T\kern-.1667em\lower.7ex\hbox{E}\kern-.125emX}

\begin{document}

\label{firstpage}

\maketitle

\begin{abstract}

Physics of the gravitational effect of the galactic bar and spiral structure is presented.
Physical equations differ from the conventionally used equations.

Application to the motion of the Sun is treated.
The speed of the Sun is taken to be consistent with the Oort constants.

Galactic radial migration of the Sun is less than $\pm$ 0.4 $kpc$ for the four-armed spiral structure. The Sun remains about 75 \% of its existence within
galactocentric distances (7.8 $-$ 8.2) $kpc$ and the results are practically independent
on the spiral structure strength. Thus, the radial distance changes only within 5\%
from the value of 8 $kpc$.

Galactic radial migration of the Sun is less than $\pm$ (0.3 $-$ 1.2) $kpc$,
for the two-armed spiral structure. The Sun remains (29 $-$ 95)\% of its existence
within galactocentric distances (7.8 $-$ 8.2) $kpc$ and the results strongly depend
on the spiral structure strength and the angular speed of the spiral arms.
The radial distance changes within (3.8 $-$ 15.0)\% from the value of 8 $kpc$.

If observational arguments prefer relevant radial migration of the Sun, then
the Milky Way is characterized by the two-arm spiral structure.

\end{abstract}


\section{Introduction}
Orbital evolution of the Sun in our galaxy, Milky Way, is important
for understanding of galactic evolution, its dynamics,
kinematics and chemical composition. It is also relevant for understanding
of evolution of the Solar System,
e.g. the effect of the galactic tides on the evolution of the \"{O}pik - Oort cloud
of comets. Several papers on the gravitational effect of the galactic
bar on motion of the Sun in the Galaxy appeared recently, e.g.
Minchev \& Famaey (2010), Famaey \& Minchev (2010), Minchev et al. (2010),
Minchev et al. (2011). One of the important results of the papers shows
that the Sun can radially migrate in the disk of the Galaxy.

Minchev \& Famaey (2010), Famaey \& Minchev (2010) present the change of the
solar galactocentric distance. The distance may change within several kiloparsecs.


If we are interested in real galactic radial migration of the Sun,
we have to treat physical equation of motion. The first aim of this paper
is to find the equation of motion, to put the relevant equations on a
firm physical basis. The second aim of this paper is to use the physical equations
in performing calculations similar to Minchev \& Famaey (2010).
We are interested in radial motion (radial migration) of the Sun in the Galaxy.

Section 2 presents multipole expansion of the galactic potential
and the concentration is paid to the quadrupole term, important for the galactic bar.
The relevant quantities found in the monopole and quadrupole terms
are discussed in Section 3, where the values of the quantities
are calculated for several models of the galactic bar. Section 4 offers the
equation of motion for an object under the gravitational influence of
the galactic disk, halo and the monopole and quadrupole terms
of the galactic bar (the monopole term corresponds to the conventional idea of the
galactic bulge). Section 5 presents numerical results for the radial migration of the Sun.
Finally, discussion and conclusion can be found in Sections 6 and 7.

\section{Multipole expansion}
Having a mass density distribution $\varrho ( \vec{r} )$, the
gravitational potential at a point $\vec{r}$ is
\begin{equation}\label{phi-total}
\Phi ( \vec{r} ) = -~G ~\int ~\frac{\varrho ( \vec{r} ~')}{
| \vec{r} - \vec{r} ~' |} ~d^{3} \vec{r} ~' ~,
\end{equation}
where $G$ is the gravitational constant.
Making an expansion in terms of the ratio
$|\vec{r} ~'| / | \vec{r} |$ $\equiv$ $r' / r$, we can write
\begin{equation}\label{phi-total-exp}
\Phi ( \vec{r} ) = \sum_{j=0}^{\infty} \Phi ^{ ( j )} ( \vec{r} )  ~,
\end{equation}
and,
\begin{equation}\label{phi0}
\Phi ^{( 0 )} ( \vec{r} ) = -~ G ~\frac{1}{r} ~
\int ~\varrho ( \vec{r} ~') ~d^{3} \vec{r} ~' = - ~\frac{G ~M}{r} ~,
\end{equation}
where $M$ is the total mass concentrated in a small region so that
the condition $|\vec{r} ~'| / | \vec{r} | \ll 1$ is fulfilled.
Other terms of the expansion are $\Phi ^{( 1 )} ( \vec{r} ) = 0$, and, the
quadrupole term is (Einstein's summation convention is used)
\begin{eqnarray}\label{phi2}
\Phi ^{( 2 )} ( \vec{r} ) &=& -~\frac{G}{2 ~r^{3}} ~Q_{ij} ~n_{i} ~ n_{j}  ~,
\nonumber \\
Q_{ij} &=& \int ~[ 3 ( \vec{r} ~')_{i} ~ ( \vec{r} ~')_{j}
~-~ | \vec{r} ~'| ^{2} \delta_{ij} ] ~\varrho ( \vec{r} ~')
~d^{3} \vec{r} ~' ~,
\nonumber \\
n_{i} &=& ( \vec{r} )_{i} / r
\end{eqnarray}
and $\delta_{ij}$ is the Kronecker delta. Since $\Phi ^{( 2 )} ( \vec{r} )$
is given by a contraction of two tensors of the second order, $Q_{ij}$
and $n_{i} ~ n_{j}$, it is an invariant independent on the coordinate basis.
The numerical calculations can be done in a special coordinate basis,
the primed coordinates, e.g., defined by the relations describing rotation
\begin{eqnarray}\label{rot}
x' &=& +~x ~\cos \alpha ~+~ y~ \sin \alpha ~,
\nonumber \\
y' &=& -~x ~\sin \alpha ~+~ y~ \cos \alpha ~,
\nonumber \\
z' &=& z ~.
\end{eqnarray}
In the case of a galactic bar we will use
\begin{equation}\label{alpha}
\alpha = \Omega_{b} ~t ~+~ \alpha_{0} ~,
\end{equation}
where $| \Omega_{b} |$ is the magnitude of the angular velocity of the bar's rotation.
In reality, $\Omega_{b}$ is negative and the negative sign denotes the observational fact that
rotation is negative, i.e., clockwise.

Eqs. (\ref{phi2})-(\ref{alpha}) yield
\begin{eqnarray}\label{kvadr}
\Phi^{(2)}(x,y,z) &=& -~\frac{G}{2~r^3} ~ \times ~X_{\Phi} ~,
\nonumber \\
X_{\Phi} &=& Q_{11} n_{1}^{2} +
Q_{22} n_{2}^{2} - \left( Q_{11} + Q_{22} \right) n_{3}^{2}   ~,
\end{eqnarray}
where
primes above the $Q_{11}$ and $Q_{22}$ terms are omitted, for the purpose of brevity.
The coordinate system is chosen in the
way that $Q_{ij}$ $=$ 0 if $i \ne j$ and some symmetry exists.
The mass density is an even function of coordinate arguments,
\begin{eqnarray}\label{Q}
Q_{11} &=& \int\limits_{-\infty}^{\infty}\rho(x,y,z)\left(2x^2-y^2-z^2\right)\,dxdydz ~,
\nonumber\\
Q_{22} &=& \int\limits_{-\infty}^{\infty}\rho(x,y,z)\left(2y^2-x^2-z^2\right)\,dxdydz
\end{eqnarray}
and
\begin{eqnarray}\label{n}
n_1^2 &=& \left( x^2 \cos^2{\alpha} + y^2 \sin^2{\alpha} + 2 x y \sin{\alpha} \cos{\alpha} \right ) / r^2 ~,
\nonumber\\
n_2^2 &=& \left ( x^2 \sin^2{\alpha} + y^2 \cos^2{\alpha} - 2 x y \sin{\alpha} \cos{\alpha} \right ) / r^2 ~,
\nonumber \\
n_3^2 &=& z^2/r^2 ~.
\end{eqnarray}
Inserting Eqs. (\ref{n}) into Eq. (\ref{kvadr}) one obtains
\begin{eqnarray}\label{kvadr1}
\Phi^{(2)}(x,y,z) &=& -~
\frac{G~Q_{11} \left ( 1 - Q_{22} / Q_{11} \right )}{4~ r^5} \times ~X_{\Phi 2}
\nonumber \\
X_{\Phi 2} &=& \frac{1 + Q_{22} / Q_{11}}{1 - Q_{22} / Q_{11}}
\left ( x^2 + y^2 - 2~z^2 \right )
\nonumber \\
& & +~\left ( x^2 - y^2 \right ) \cos{(2 \alpha)}
~+~2 ~x~ y~ \sin{(2 \alpha)}  ~,
\nonumber \\
\alpha &=& \Omega_{b} ~t ~+~ \alpha_{0} ~,
\end{eqnarray}
if also Eq. (\ref{alpha}) is added.
As for the Milky Way, the value of $\alpha_{0}$ is about $-$25 degrees
(compare with Majaess 2010; the real value may differ in more than 10 degrees)
and the value of the angular velocity is $\Omega_{b}$ $=$ $-~55.5~km~s^{-1}~kpc^{-1}$
(compare with Minchev \& Famaey 2010), $t$ is the time.

\subsection{Discussion on Eq. (\ref{kvadr1})}
Our result represented by Eq.  (\ref{kvadr1}) differs from the result
conventionally used by other authors. The conventional result is
(see, e.g., Minchev \& Famaey 2010, Dehnen 2000)
\begin{eqnarray}\label{kvadr1-conv}
\Phi^{(2)}_{conv} &=& \frac{1}{2}~ Q_{T} ~v_{c}^{2} ~
\cos{[2 (\phi - \Omega_{b} t)]} \times X_{\Phi 2 c} ~,
\nonumber \\
 X_{\Phi 2 c} &=& \left ( \frac{r_{b}}{r} \right )^{3} ~,
~~r \ge r_{b} ~,
\nonumber \\
 X_{\Phi 2 c} &=& 2 - \left ( \frac{r}{r_{b}} \right )^{3}  ~, ~~r \le r_{b} ~,
\end{eqnarray}
where $r_{b} = 3.44~kpc$, $v_{c} = 240~km~s^{-1}$, $0.1 < Q_{T} < 0.4$
(Minchev \& Famaey 2010).

At first, the second formula of Eqs. (\ref{kvadr1-conv}) does not hold. The inner
part of the galactic bar potential must be calculated from the Poisson
equation or its equivalence: $\bigtriangleup \Phi = 4 \pi G \varrho$,
$\Phi = -~G ~\int ~ \varrho ( \vec{r} ~') /
| \vec{r} - \vec{r} ~' | ~d^{3} \vec{r} ~'$. As for the case $r \ge r_{b}$
in Eqs. (\ref{kvadr1-conv}), it leads to
\begin{eqnarray}\label{kvadr1-c1}
\Phi^{(2)}_{conv}(x,y,z) &=& \frac{1}{2}~ Q_{T} ~v_{c}^{2} ~
\left ( \frac{r_{b}}{\sqrt{R^{2} + z^{2}}} \right )^{3} ~\frac{1}{R^2}
~\times  Y_{\Phi 2 c}
\nonumber \\
Y_{\Phi 2 c} &=& \left ( x^2 - y^2 \right ) ~
\cos ( 2 \Omega_{b} t )
\nonumber \\
& & +~ 2 ~x~ y~ \sin ( 2 \Omega_{b} t)  ~.
\end{eqnarray}
We remind that $r = \sqrt{R^{2} + z^{2}}$,
where $R$ is the 2-dimensional radial coordinate in the galactic equatorial plane
and $z$ is the vertical coordinate normal to the equatorial plane.
One can immediately see that physical Eq. (\ref{kvadr1}) differs from the conventional potential.

\section{Models of the galactic bar}
As examples of mass distribution in the galactic bar we take
the models presented by Stanek et al. (1997). For the models
G1, G3, E2, P1, P2, P3 we introduce the following
complete nonsymmetric ellipsoid coordinates ($r$ is a dimensionless quantity, now --
it differs from the quantity used in the previous section):
\begin{eqnarray}\label{complete nonsymmetric ellipsoid coordinates}
x &=& r ~x_0 ~\sin{\theta} ~\cos{\varphi} ~,
\nonumber \\
y &=& r ~y_0 ~\sin{\theta} ~\sin{\varphi} ~,
\nonumber \\
z &=& r ~z_0 ~\cos{\theta} ~,
\end{eqnarray}
with the corresponding volume element
\begin{equation}\label{volume element}
dx~dy~dz = x_0~y_0~z_0 ~r^2 ~\sin{\theta} ~d\theta ~d\varphi ~dr \, .
\end{equation}
The corresponding quantities $Q_{11}$ and $Q_{22}$, and, also mass of the galactic bar $M$,
can now be calculated in an analytical way.

\begin{enumerate}
\item The model G1 with mass density
\begin{eqnarray}\label{hustota G1}
    &&\rho_{G1}(x,y,z)=\rho_0\exp{\left( -~ r^2 / 2 \right )} \,,
    \nonumber\\
    &&r=\sqrt{\left(\frac{x}{x_0}\right)^2 + \left(\frac{y}{y_0}\right)^2 + \left(\frac{z}{z_0}\right)^2}
  \end{eqnarray}
yields
  \begin{eqnarray}\label{Q11 a pomer pre G1}
  M &=& 2~\sqrt{2} ~\pi^{3/2}~\rho_0~x_0~y_0~z_0 ~,
   \nonumber \\
    Q_{11} &=& 2 ~\sqrt{2} ~\pi^{3/2} ~\rho_0 ~x_0 ~y_0 ~z_0 ~\left (
    2x_0^2 - y_0^2 - z_0^2 \right )
    \nonumber \\
&=& M~\left ( 2x_0^2 - y_0^2 - z_0^2 \right ) ~,
    \nonumber \\
    \frac{Q_{22}}{Q_{11}} &=& \frac{2 y_0^2 - x_0^2 - z_0^2}{2 x_0^2 - y_0^2 - z_0^2} ~.
  \end{eqnarray}
  \item The model G3 defined by the mass density
  \begin{equation}\label{hustota G3}
    \rho_{G3}(x,y,z)= \tilde{\rho}_0  ~r^{-1.8}\exp{\left(-~ r^3 \right)} ~,
  \end{equation}
produces
  \begin{eqnarray}\label{Q11 a pomer pre G3}
    M &=& 4\pi \times 0.7394 \times \tilde{\rho}_0~x_0~y_0~z_0 ~,
    \nonumber\\
    Q_{11} &\doteq& \frac{4 \pi}{3} \times 0.3219 \times \tilde{\rho}_0 ~x_0 ~y_0 ~z_0 ~ \left ( 2x_0^2 - y_0^2 -z_0^2 \right )
    \nonumber \\
    &\doteq& 0.1451\times M~\left ( 2x_0^2 - y_0^2 -z_0^2 \right )~,
    \nonumber\\
    \frac{Q_{22}}{Q_{11}} &=& \frac{2y_0^2-x_0^2-z_0^2}{2x_0^2-y_0^2-z_0^2}\,,
  \end{eqnarray}
where the number $0.3219$ is an approximate value of the integral
$$\int\limits_0^{\infty} r^{2.2} \exp{\left( -~ r^3 \right)}\,dr\,.$$
  \item The model E2 defined by the mass density
  \begin{equation}\label{hustota E2}
    \rho_{E2}(x,y,z)=\rho_0\exp{\left( -~r \right)}
  \end{equation}
gives
  \begin{eqnarray}\label{Q11 a pomer pre E2}
  M &=& 8 \pi ~\rho_0~x_0~y_0~z_0 ~,
  \nonumber\\
 Q_{11} &\doteq& 32 \pi ~\rho_0 ~x_0 ~y_0 ~z_0 \left ( 2 x_0^2 - y_0^2 -z_0^2 \right )
 \nonumber \\
 &=& 4 ~ M ~ \left ( 2 x_0^2 - y_0^2 -z_0^2 \right ) ~,
 \nonumber \\
 Q_{22}/Q_{11} &=& \frac{2y_0^2-x_0^2-z_0^2}{2x_0^2-y_0^2-z_0^2} ~.
  \end{eqnarray}
  \item The model P1 with mass density
  \begin{equation}\label{hustota P1}
    \rho_{P1}(x,y,z) = \frac{\rho_0}{\left ( 1 + r \right )^4}
  \end{equation}
yields
  \begin{eqnarray}\label{Q11 pre P1}
  M &=& 4\pi ~ \rho_0 ~ x_0 ~ y_0 ~ z_0 \int\limits_{0}^{{r}_{bar}} \frac{r^2}{\left(1+r\right)^4}\,dr
  \nonumber \\
  & & \nonumber \\
  &=& \frac{4\pi}{3} ~ \rho_0 ~ x_0 ~ y_0~z_0 ~\frac{r^3_{bar}}{\left(1+r_{bar}\right)^3}  ~,
  \nonumber\\
    Q_{11}&=&\frac{4\pi}{3} ~ \rho_0 ~x_0 ~y_0 ~z_0
    \left ( 2x_0^2 - y_0^2 - z_0^2\right) \int\limits_{0}^{r_{bar}} \frac{r^4}{\left(1+r\right)^4}\,dr
    \nonumber\\
    &=&\frac{4\pi}{3} ~ \rho_0 ~x_0 ~y_0 ~z_0
    \left ( 2x_0^2 - y_0^2 - z_0^2\right) ~\times X_{P1}
    \nonumber \\
    &=& \frac13~M\left ( 2x_0^2 - y_0^2 - z_0^2\right) \left[3r_{bar}+22+\frac{30}{r_{bar}}
    \right.
    \nonumber\\
    & & +~ \frac{12}{r^2_{bar}} ~-~ \left.12 \left(1+\frac1{r_{bar}}\right)^3\ln{\left(1+r_{bar}\right)}
    \right] ~,
    \nonumber \\
    \nonumber \\
    X_{P1} &=& \frac{3r^4_{bar}+22r^3_{bar}+30r^2_{bar}+12r_{bar}}{3\left(1+r_{bar}
    \right)^3} - 4 \ln{\left(1+r_{bar}\right)}
    \nonumber\\
    r_{bar} &\equiv& r_{b} / x_{0} ~~~(as~an~example) ~.
  \end{eqnarray}
We have taken the symbol $r_{bar}$ as the upper limit of the integral in Eq. (\ref{Q11 pre P1})
(if one would like to use infinity as the upper limit of the integral, then the integral would diverge,
$Q_{11} \rightarrow \infty$. If we take into account that the real length of the bar $r_b$
is proportional to $x_{0}$, then we can take $r_{bar}$ $=$ $r_{b} /x_{0}$.

\item The model P2 with mass density
  \begin{equation}\label{hustota P2}
    \rho_{P2}(x,y,z) = \frac{\rho_0}{r\left(1+r\right)^3}
  \end{equation}
yields
\begin{eqnarray}\label{MP2}
M &=& 4\pi ~ \rho_0 ~ x_0 ~ y_0 ~ z_0 \int\limits_{0}^{r_{bar}}\frac{r}{\left(1+r\right)^3}\,dr \nonumber\\
  & & \nonumber\\
  &=& 2 \pi ~ \rho_0 ~ x_0 ~ y_0~z_0 ~\frac{r^2_{bar}}{\left(1+r_{bar}\right)^2}  ~, \nonumber\\
Q_{11}&=&\frac{4\pi}{3} ~ \rho_0 ~x_0 ~y_0 ~z_0
    \left ( 2x_0^2 - y_0^2 - z_0^2\right) \int\limits_{0}^{r_{bar}}\frac{r^3}{\left(1+r\right)^3}\,dr
    \nonumber\\
    &=&\frac{4\pi}{3} ~ \rho_0 ~x_0 ~y_0 ~z_0
    \left ( 2x_0^2 - y_0^2 - z_0^2\right) \times X_{P2}
    \nonumber\\
    &=& \frac13~M\left ( 2x_0^2 - y_0^2 - z_0^2\right)
    \left[2r_{bar}+9+\frac{6}{r_{bar}} \right .
    \nonumber \\
    & & \left .
    -~ 6 \left(1+\frac1{r_{bar}}\right)^2\ln{\left(1+r_{bar}\right)}\right] ~,
    \nonumber \\
    \nonumber \\
    X_{P2} &=& \frac{2r^3_{bar}+9r^2_{bar}+6r_{bar}}{2
    \left(1+r_{bar}\right)^2} - 3\ln{\left(1+r_{bar}\right)}
    \nonumber\\
    r_{bar} &\equiv& r_{b} / x_{0} ~~~(as~an~example) ~.
\end{eqnarray}
\item The model P3 with mass density
  \begin{equation}\label{hustota P3}
    \rho_{P3}(x,y,z) = \frac{\rho_0}{\left( 1 + r^2 \right)^2}
  \end{equation}
yields
\begin{eqnarray}\label{MP3}
M &=& 4\pi ~ \rho_0 ~ x_0 ~ y_0 ~ z_0 \int\limits_{0}^{r_{bar}}
\frac{r^{2}}{\left ( 1 + r^2 \right) ^2} \,dr
\nonumber \\
&=& 2\pi ~ \rho_0 ~ x_0 ~ y_0 ~ z_0 ~
\left( \arctan{r_{bar}} - \frac{r_{bar}}{1+r^2_{bar}} \right) ~,
\nonumber\\
Q_{11}&=&\frac{4\pi}{3} ~ \rho_0 ~x_0 y_0 z_0
    \left ( 2x_0^2 - y_0^2 - z_0^2 \right) \int\limits_{0}^{r_{bar}}
    \frac{r^{4}}{\left ( 1 + r^2 \right) ^2} \,dr
    \nonumber\\
    &=& \frac{4\pi}{3} ~ \rho_0 ~x_0 ~y_0 ~z_0
    \left ( 2x_0^2 - y_0^2 - z_0^2\right) ~\times
    \nonumber \\
    & & \left(r_{bar}+\frac12\frac{r_{bar}}{1+r^2_{bar}}-\frac32\arctan{r_{bar}}\right)
    \nonumber \\
    &=&\frac23 ~M~\left ( 2x_0^2 - y_0^2 - z_0^2\right) ~\times
    \nonumber\\
    & & \frac{r_{bar}+r_{bar}/ \left [ 2 \left ( 1 + r^2_{bar} \right ) \right ] -(3/2)\arctan{r_{bar}}}{
    \arctan{r_{bar}} - r_{bar}/ \left(1+r^2_{bar}\right)} ~,
    \nonumber \\
    \nonumber \\
    r_{bar} &\equiv& r_{b} / x_{0} ~~~(as~an~example)~.
\end{eqnarray}
\end{enumerate}

As for the model E1 with mass density
\begin{equation}\label{hustota E1}
    \rho_{E1}(x,y,z)=\rho_0 \exp{(-r_e)}\,,
\end{equation}
where
\begin{equation}\label{re}
    r_e=\frac{|x|}{x_0}+\frac{|y|}{y_0}+\frac{|z|}{z_0}\, ,
\end{equation}
the values of $M$, $Q_{11}$ a $Q_{22}$ can be easily analytically calculated
(one may avoid Eqs. \ref{complete nonsymmetric ellipsoid coordinates} --
\ref{volume element}):
\begin{eqnarray}\label{Q11 a pomer pre E1}
    M &=& 8 ~\rho_0~x_0~y_0~z_0 ~,
    \nonumber\\
    Q_{11} &=& 16 ~\rho_0x_0y_0z_0\left(2x_0^2 - y_0^2 - z_0^2\right)
    \nonumber\\
    &=& 2M~ \left(2x_0^2 - y_0^2 - z_0^2\right)~,
    \nonumber\\
    Q_{22}/Q_{11} &=& \frac{2y_0^2 - x_0^2 - z_0^2}{2x_0^2 - y_0^2 - z_0^2} ~.
\end{eqnarray}

The model G2 with mass density
\begin{equation}\label{hustota G2}
    \rho_{G2}(x,y,z)=\rho_0\exp{\left(-r_{s}^2/2\right)} ~,
  \end{equation}
  where
  \begin{equation}\label{rs}
    r_s=\left\{\left[\left(\frac x{x_0}\right)^2 + \left(\frac y{y_0}\right)^2\right]^2+\left(\frac z{z_0}\right)^4\right\}^{1/4} ~,
  \end{equation}
can be solved with the following substitutions:
\begin{eqnarray}\label{G2-RS-upperhemisphere}
x &=& x_{0} ~r_s ~\sin ^{1/2} \vartheta ~\cos \varphi ~,
\nonumber \\
y &=& y_{0} ~r_s ~\sin ^{1/2} \vartheta ~\sin \varphi ~,
\nonumber \\
z &=& z_{0} ~r_s ~\cos ^{1/2} \vartheta ~,
\end{eqnarray}
for the upper hemisphere, and,
\begin{eqnarray}\label{G2-RS-lowerhemisphere}
x &=& x_{0} ~r_s ~\sin ^{1/2} \vartheta ~\cos \varphi ~,
\nonumber \\
y &=& y_{0} ~r_s ~\sin ^{1/2} \vartheta ~\sin \varphi ~,
\nonumber \\
z &=& -~z_{0} ~r_s ~\cos ^{1/2} \vartheta ~,
\end{eqnarray}
for the lower hemisphere, and, for both hemispheres
$r_s$ $\in$ $\langle 0, \infty )$,
$\vartheta$ $\in$ $\langle 0, \pi / 2 \rangle$,
$\varphi$ $\in$ $\langle 0, 2 \pi )$, and, the volume element is
\begin{equation}\label{specialne sfericke suradnice}
dx~dy~dz =\frac12~x_0~y_0~z_0 ~r_s^2 ~\cos^{-1/2}{\theta} ~d\theta ~d\varphi ~dr_s
\end{equation}
for both hemispheres. The values of the quantities $M$, $Q_{11}$ and $Q_{22}/Q_{11}$ are:
  \begin{eqnarray}\label{Q11 a pomer pre G2}
    M &=& 2 ~\pi^{3/2} ~K\left(\frac12\right)~\rho_0~x_0~y_0~z_0 ~,
    \nonumber\\
    Q_{11} &=& 6\pi^{3/2}\sqrt{2}~\rho_0 ~x_0 ~y_0 ~z_0 \times X_{G2}
    \nonumber \\
    &=& 3\sqrt{2} \left[K\left(\frac12\right)\right]^{-1}M \times X_{G2} \,,
    \nonumber \\
    \frac{Q_{22}}{Q_{11}} &=& \frac{2y_0^2 - x_0^2 -
    \sqrt{2 / \pi} \left [ \Gamma \left (3 / 4 \right) \right ]^2
    z_0^2}{2x_0^2 - y_0^2 -
    \sqrt{2 / \pi} \left [ \Gamma \left (3 / 4 \right) \right ]^2 z_0^2} ~\,,
    \nonumber\\
    X_{G2} &=& 2x_0^2 - y_0^2 -\sqrt{\frac2{\pi}} \left [
    \Gamma\left(\frac34\right) \right ]^2 z_0^2 ~,
  \end{eqnarray}
where $K(x)$ is the complete elliptic integral of the first kind.

Finally, the model E3
\begin{equation}\label{hustota E3}
    \rho_{E3}(x,y,z)=\rho_0 ~K_0(r_s) ~,
  \end{equation}
yields
\begin{eqnarray}\label{Q11 a pomer pre E3}
M &=& \sqrt{2} ~\pi^2 ~K \left(\frac12\right)~\rho_0~x_0~y_0~z_0 ~,
\nonumber\\
Q_{11} &=& 18~\pi^{2}~\rho_0 ~x_0 ~y_0 ~z_0 \times X_{E3}
\nonumber \\
&=& 9\sqrt{2}\left[K\left(\frac12\right)\right]^{-1}M \times X_{E3} ~,
\nonumber \\
    \frac{Q_{22}}{Q_{11}} &=& \frac{2y_0^2 - x_0^2 -\sqrt{2 / \pi}
    \left [ \Gamma \left (3 / 4 \right) \right ]^2 z_0^2}{2x_0^2 - y_0^2 -
    \sqrt{2 / \pi} \left [ \Gamma \left (3 / 4 \right) \right ]^2  z_0^2} ~,
    \nonumber\\
    X_{E3} &=& 2x_0^2 - y_0^2 -\sqrt{\frac2{\pi}} \left [
    \Gamma\left(\frac34\right) \right ]^2 z_0^2 ~\, ~.
  \end{eqnarray}

\begin{table}
\small
\begin{center}
\caption{Numerical values of the quantities discussed in the previous subsection.
Data for $x_{0}$,  $y_{0}$ and $z_{0}$ are taken from Stanek et al. (1997).
The quantity $M/ \rho_{0}$ contains the central density $\rho_{0}$.}
\begin{tabular}{|l||r|r|r|}
  \hline
  \multicolumn{1}{|c||}{Model} & \multicolumn{1}{|c|}{$x_0$} & \multicolumn{1}{|c|}{$y_0$} & \multicolumn{1}{|c|}{$z_0$} \\
  & \multicolumn{1}{|c|}{$[pc]$} & \multicolumn{1}{|c|}{$[pc]$} & \multicolumn{1}{|c|}{$[pc]$}
  \\ \hline
  G1  & $1478$ & $754$  & $611$ \\
  G3  & $4537$ & $2303$ & $1725$ \\
  E2$^{(1)}$  & $1000$ & $500$  & $400$ \\
  E2$^{(2)}$ & $1000$ & $505$  & $388$ \\
  P1  & $1565$ & $783$  & $582$ \\
  P2  & $2653$ & $1320$ & $975$ \\
  P3  & $1869$ & $939$  & $713$ \\
  E1  & $1650$ & $868$  & $474$ \\
  G2  & $1298$ & $694$  & $572$ \\
  E3  & $1075$ & $563$  & $442$ \\
  \hline
\end{tabular}
\begin{tabular}{|l||c|c|c|}
  \hline
  \multicolumn{1}{|c||}{Model}
  & \multicolumn{1}{|c|}{$Q_{22}/Q_{11}$} & \multicolumn{1}{|c|}{$Q_{11}/M$}
  & \multicolumn{1}{|c|}{$M/ \rho_{0}$} \\
  & \multicolumn{1}{|c|}{[ -- ]} &
  \multicolumn{1}{|c|}{[ $pc^{2}$ ]} & \multicolumn{1}{|c|}{[ $pc^{3}$ ]}
  \\ \hline
  G1  & $-0.4146$ & $3.4271\times10^6$ & $1.0724\times10^{10}$ \\
  G3  & $-0.3938$ & $4.7727\times10^6$ & -- \\
  E2$^{(1)}$  & $-0.4151$ & $6.3600\times10^6$ &
  $5.0266\times10^{9}$ \\
  E2$^{(2)}$ & $-0.4017$ & $6.3777\times10^6$ &
  $4.9245\times10^{9}$ \\
  P1  & $-0.3957$ & -- & -- \\
  P2  & $-0.3957$ & -- & -- \\
  P3  & $-0.3999$ & -- & -- \\
  E1  & $-0.3224$ & $8.9338\times10^6$ &
  $5.4309\times10^{9}$ \\
  G2  & $-0.4461$ & $5.7115\times10^6$ &
  $1.0639\times10^{10}$ \\
  E3  & $-0.4294$ & $1.2084\times10^7$ &
  $6.9228\times10^{9}$ \\
  \hline
\end{tabular}
\end{center}
\end{table}

\subsection{Numerical results}

Table 1 presents numerical results for the data given by
Stanek et al. (1997 -- Table 4). Since the integrations in the models
P1, P2 and P3 lead to integrals which require finite limits in order
to be convergent, we have considered the upper limit corresponding to
the length of the bar $r_b$ and the values of $x_0$ for individual models
(see Eqs. \ref{Q11 pre P1}, \ref{MP2}, \ref{MP3}). Taking into account the value $r_b =$ 3.44 $kpc$,
the required quantities are presented in Table 2. If we admit a relative error of $r_b$ to be 20 \%,
then the relative errors of the quantities are also presented in Table 2.

\begin{table}
\small
\begin{center}
\caption{Numerical values of quantities for the models P1, P2 and P3.
The values of $Q_{11}/M$ and $M/ \rho_{0}$
hold for $r_b =$ 3.44 $kpc$. The relative errors
$\varrho (Q_{11}/M)$ and $\varrho (M/ \rho_{0})$
hold for the uncertainty of 20 \% in $r_b$.}
\begin{tabular}{|l||c|c|c|}
  \hline
  \multicolumn{1}{|c||}{Model} &
  \multicolumn{1}{|c|}{$Q_{22}/Q_{11}$}
  & \multicolumn{1}{|c|}{$Q_{11}/M$}
  & \multicolumn{1}{|c|}{$\varrho (Q_{11}/M)$} \\
  & \multicolumn{1}{|c|}{[ -- ]} &
  \multicolumn{1}{|c|}{[ $pc^{2}$ ]} & \multicolumn{1}{|c|}{[ \% ]} \\
  \hline
P1 & $-0.3957$ & $2.3149 \times 10^6$ & $-31 ~/~+34$ \\
P2 & $-0.3957$ & $2.1628 \times 10^6$ & $-32 ~/~+36$ \\
P3 & $-0.3999$ & $2.5152 \times 10^6$ & $-29 ~/~+31$  \\
\hline
\end{tabular}
\begin{tabular}{|l||c|c|c|c|c|}
  \hline
  \multicolumn{1}{|c||}{Model} &
  \multicolumn{1}{|c|}{$Q_{22}/Q_{11}$}
  & \multicolumn{1}{|c|}{$M/ \rho_{0}$}
  & \multicolumn{1}{|c|}{$\varrho (M/ \rho_{0})$} \\
  & \multicolumn{1}{|c|}{[ -- ]} &
  \multicolumn{1}{|c|}{[ $pc^{3}$ ]} & \multicolumn{1}{|c|}{[ \% ]}
  \\ \hline
P1 & $-0.3957$ & $0.9700 \times 10^9$ & $-20 ~/~+17$ \\
P2 & $-0.3957$ & $6.8383 \times 10^9$ & $-19 ~/~+16$ \\
P3 & $-0.3999$ & $5.1389 \times 10^9$ & $-22 ~/~+18$ \\
\hline
\end{tabular}
\end{center}
\end{table}

\section{Conventional and physical approaches}
In order to compare the published results with our approach, we have
to summarize important equations.

\subsection{Conventional approach}
The conventional approach to galactic motion of the Sun defines the
following potentials of the Galaxy (Minchev et al. 2010).
The background axisymmetric potential due to the disk and halo has the form
\begin{eqnarray}\label{conv-disk+halo}
\Phi_{0~conv} &=& v_{c}^{2} ~\ln ( r / r_{0} )
\end{eqnarray}
corresponding to a flat rotation curve. The non-axisymmetric potential due to
the galactic bar is modeled as a pure quadrupole
\begin{eqnarray}\label{conv-bar}
\Phi^{(2)}_{b~conv} &=& \frac{1}{2}~ Q_{T} ~v_{c}^{2} ~
\cos{[2 (\phi - \Omega_{b} t)]} \times X_{\Phi^{(2)}_{b~conv}} ~,
\nonumber \\
X_{\Phi^{(2)}_{b~conv}} &=& \left ( \frac{r_{b}}{r} \right )^{3} ~, ~~r \ge r_{b} ~,
\nonumber \\
X_{\Phi^{(2)}_{b~conv}} &=&
\left [ 2 - \left ( \frac{r}{r_{b}} \right )^{3} \right ] ~, ~~r \le r_{b} ~,
\end{eqnarray}
where $r_{b} = 3.44~kpc$, $v_{c} = 240~km~s^{-1}$, $0.1 < Q_{T} < 0.4$
(see also Minchev \& Famaey 2010) and the spiral potential is given by
\begin{eqnarray}\label{conv-spiral arms}
\Phi_{s~conv} &=& \epsilon_{s} ~
\cos{ \left [ \tilde{\alpha} ~\ln \left ( r / r_{0} \right ) ~-~ m
( \phi - \Omega_{s} t) \right ]} ~,
\end{eqnarray}
where $\epsilon_{s}$ is the spiral structure strength,
0.015 $<$ $| \epsilon_{s} |$ $<$ 0.072 for a two-armed spiral structure and
0.007 $<$ $| \epsilon_{s} |$ $<$ 0.031 for a four-armed spiral structure,
$\epsilon_{s}$ $=$ $sign (-~\tilde{\alpha})$ $\times$ $| \epsilon_{s} |$,
$\tilde{\alpha}$ $=$ $-4$ and $-8$ for the two-armed and four-armed spiral structure,
respectively, where the negative sign corresponds to trailing spirals,
$m$ is an integer corresponding to the number of arms,
$\Omega_{s}$ $\in$ $\left \{ 0.7, 0.8, 0.9, 1.0, 1.1 \right \} \times
v_{c} / R_{0}$, i.e.,
$\Omega_{s} [km~s^{-1} ~kpc^{-1}]$ $\in$ $\left \{ 21, 24, 27, 30, 33
\right \}$, if $v_{c}$ $=$ 240 $km~s^{-1}$ and $R_{0}$ $=$ 8.0 $kpc$.

\subsection{Physical approach}
In this section we take the background axisymmetric potential due to the
disk and halo in the form similar to Eq. (\ref{conv-disk+halo})
\begin{eqnarray}\label{phys-disk+halo}
\Phi_{disk+halo} &=& -~v_{c}^{2} ~
\left \{ 1 ~-~ \ln ( r / r_{G} ) \right \} ~,
\end{eqnarray}
where $r_{G}$ is the dark matter radius of the Galaxy. Eq. (\ref{phys-disk+halo})
is the physical potential corresponding to the flat rotation curve
defined by the condition $v(r)$ $=$ $v_{c}$. The result presented by
Minchev \& Famaey (2010) or Dehnen (2000), see our Eq. (\ref{conv-disk+halo}),
differs from Eq. (\ref{phys-disk+halo}).

The potential due to the galactic bar is given by Eqs. (\ref{phi0}) and
(\ref{kvadr1}):
\begin{eqnarray}\label{bar-phi0+kvadr10}
\Phi_{bar} ( \vec{r} ) &=& \Phi ^{( 0 )} ( \vec{r} ) ~+~
\Phi ^{( 2 )} ( \vec{r} ) ~,
\nonumber \\
\Phi ^{( 0 )} ( \vec{r} ) &=& -~ G ~\frac{1}{r} ~
\int ~\varrho ( \vec{r} ~') ~d^{3} \vec{r} ~' = - ~\frac{G ~M}{r} ~,
\nonumber \\
\Phi^{(2)}(x,y,z) &=& -~
\frac{G~Q_{11} \left ( 1 - Q_{22} / Q_{11} \right )}{4~ r^5}
\nonumber\\
& & \times \left [ \frac{1 + Q_{22} / Q_{11}}{1 - Q_{22} / Q_{11}}
\left ( x^2 + y^2 - 2~z^2 \right )  \right.
\nonumber \\
& & \left. +~\left ( x^2 - y^2 \right ) \cos{(2 \alpha)}
+ 2 x y \sin{(2 \alpha)} \right ] ~,
\nonumber \\
M &=& 1.4 \times 10^{10} M_{\odot} ~,
\nonumber \\
Q_{11} [ M_{\odot}~pc^{2} ] &=& 8 \times 10^{6} ~\times M [ M_{\odot} ] ~,
\nonumber \\
Q_{22} / Q_{11} &=& -~2/5 ~,
\nonumber \\
\alpha &=& \Omega_{b} ~t ~+~ \alpha_{0} ~,
\nonumber \\
\alpha_{0} &\doteq& -~25^{\circ}~,
\nonumber \\
\Omega_{b} &=& -~55.5~km~s^{-1}~kpc^{-1}~,
\end{eqnarray}
where also the value of $M$ from Dauphole \& Colin (1995 - Table 5) is used
and the values from Table 1 are taken into account.


The spiral potential is given by the conventional model represented by
Eq. (\ref{conv-spiral arms}), in a slightly corrected form ($R_{b}$ $=$
3.44 $kpc$):
\begin{eqnarray}\label{corr-spiral arms}
\Phi_{s~corr} &=& \epsilon_{s} v_{c}^{2}
\cos{ \left [ \tilde{\alpha} ~\ln \left ( ~R / R_{b} \right ) ~-~ m
( \phi - \phi_{s0} + \Omega_{s} t) \right ]}
\nonumber \\
&=& \epsilon_{s} v_{c}^{2}  \left \{ S_{1}
\cos{\left (m \phi_{s0} + \tilde{\alpha} \ln \frac{R}{R_{b}} \right )} \right. \nonumber\\
& & \left . ~+~S_{2} \sin{\left (m \phi_{s0} + \tilde{\alpha} \ln \frac{R}{R_{b}} \right )}\right \} ~,
\nonumber \\
S_{1} &=& \cos{\left ( m \phi \right )} \cos{\left ( m \Omega_{s} t \right )} -
\sin{\left ( m \phi \right )} \sin{\left ( m \Omega_{s} t \right )} ~,
\nonumber \\
S_{2} &=& \sin{\left ( m \phi \right )} \cos{\left ( m \Omega_{s} t \right )} +
\cos{\left ( m \phi \right )} \sin{\left ( m \Omega_{s} t \right )} ~,
\nonumber \\
\cos{\left ( 4 \phi \right )} &=& 4~ \left [ \left ( \frac{x}{R} \right ) ^{4} ~+~
\left ( \frac{y}{R} \right ) ^{4} \right ] ~-~ 3 ~,
\nonumber \\
\sin{\left ( 4 \phi \right )} &=& 4~ \frac{x}{R} ~ \frac{y}{R} ~\left [
\left ( \frac{x}{R} \right ) ^{2} ~-~ \left ( \frac{y}{R} \right ) ^{2}
\right ] ~,
\nonumber \\
\cos{\left ( 2 \phi \right )} &=& \left ( \frac{x}{R} \right ) ^{2} ~-~
\left ( \frac{y}{R} \right ) ^{2} ~,
\nonumber \\
\sin{\left ( 2 \phi \right )} &=& 2~\frac{x}{R} ~\frac{y}{R} ~,
\nonumber \\
\nonumber \\
R &=& \sqrt{x^{2} + y^{2}} ~,
\nonumber \\
\phi_{s0} &=& \alpha_{0} ~,
\end{eqnarray}
where the quantity $\phi_{s0}$ denotes orientation
of spiral arms in the coordinates x-y at the intial time $t$ $=$ 0
and we put $\phi_{s0}$ $=$ $\alpha_{0}$. The quantity
$\epsilon_{s}$ is the spiral structure strength,
0.015 $<$ $| \epsilon_{s} |$ $<$ 0.072 for a two-armed spiral structure and
0.007 $<$ $| \epsilon_{s} |$ $<$ 0.031 for a four-armed spiral structure,
$\epsilon_{s}$ $=$ $sign (\tilde{\alpha})$ $\times$ $| \epsilon_{s} |$,
$\tilde{\alpha}$ $=$ $+~4$ and $+~8$ for the two-armed and four-armed spiral structure,
respectively, where the positive sign corresponds to trailing spirals,
$m$ is an integer corresponding to the number of arms,
$\Omega_{s}$ $\in$ $\left \{ 0.7, 0.8, 0.9, 1.0, 1.1 \right \}$ $\times$
$v_{y0} / R_{0}$, where $v_{y0}$ is the initial circular velocity
of the Sun and $R_{0}$ $=$ 8.0 $kpc$.

\subsection{Equation of motion}

If an object moves in the Galaxy under the action of galactic disk, halo, galactic bar
and spiral structure, then the equation of motion of the object reads
\begin{eqnarray}\label{eqm-phys}
\dot{\vec{v}} &=& -~ grad \left ( \Phi_{disk+halo} ~+~  \Phi_{bar} ~+~  \Phi_{s~corr} \right ) ~,
\end{eqnarray}
where Eqs. (\ref{phys-disk+halo}), (\ref{bar-phi0+kvadr10}) and (\ref{corr-spiral arms}) can be used.

\subsection{Motion of the Sun -- initial conditions}

Initial conditions for the Sun are
\begin{eqnarray}\label{Sun-motion-init.c.}
x &=& R_{0} = 8~kpc~,
\nonumber \\
y &=& 0~,
\nonumber \\
v_{x} &=& 0~,
\nonumber \\
v_{y} &=& v_{y0} = -~255.2~km~s^{-1}~,
\nonumber \\
& & for~Eqs.~ (\ref{phys-disk+halo},~
\ref{bar-phi0+kvadr10},~ \ref{corr-spiral arms}) ~-~ the~first~case ~,
\nonumber \\
v_{y} &=& v_{y0} = -~220.0~km~s^{-1}~,
\nonumber \\
& &  for~Eqs.~ (\ref{phys-disk+halo},~
\ref{bar-phi0+kvadr10},~ \ref{corr-spiral arms}) ~-~ the~second~case ~,
\nonumber \\
v_{y} &=& v_{y0} = +~ 240.0~km~s^{-1}~,
\nonumber \\
& & for~Eqs.~ (\ref{conv-disk+halo},~
\ref{conv-bar},~ \ref{conv-spiral arms}) ~,
\end{eqnarray}
where the value of $v_y$ follows from the relations
\begin{eqnarray}
  v_y &=& -~ \sqrt{v_c^2 + \Sigma(v^2)}~,
  \nonumber \\
  \Sigma(v^2) &=& \left( \left [ v(R_0) \right ]^2 \right )_{\Phi^{(0)}} +
  \left( \left [ v(R_0) \right ]^2 \right )_{\Phi^{(2)}} ~,
  \nonumber \\
  \left(\left[v(R_0)\right]^2\right)_{\Phi^{(0)}} &=&
  \left [ R~\frac{\partial \Phi^{(0)}}{\partial R}
  \right ]_{R_0} ~=~ \frac{GM}{R_0} ~,
  \nonumber \\
  \left(\left [ v(R_0) \right ] ^2\right)_{\Phi^{(2)}} &=&
  \left[ R~ \frac{\partial}{\partial R} \left \{
  -~\frac{G~Q_{11}}{4R^3} \left ( 1 + \frac{Q_{22}}{Q_{11}} \right ) \right \}
  \right ]_{R_0}
  \nonumber \\
  &=& \frac3{32}~ \left ( 1 + \frac{Q_{22}}{Q_{11}} \right )
  \times \frac{GM}{R_0}~,
\end{eqnarray}
for the physical approach.
Eqs. (\ref{conv-disk+halo})-(\ref{Sun-motion-init.c.}) uniquely determine motion of the Sun in the Galaxy.

\section{Computational results}

We numerically solved equation of motion in accordance with Eq. (\ref{eqm-phys}) and the results
presented in the previous section.
We used both of the physical values of $v_{y}$, as it is given by Eqs. (\ref{Sun-motion-init.c.}
-- the first and the second cases).

\subsection{Effect of the galactic bar}

Table 3 shows the extremal values of the evolution of the galactocentric distance of the Sun
under the action of galactic halo, disk and galactic bar. The results presented in Table 3
hold for the special case when the effect of spiral structure is ignored.
While the nonexistence of the nonsymmetric parts in term $\Phi^{(2)}$ would produce circular orbit
for the initial conditions given by Eqs. (\ref{Sun-motion-init.c.} - the first and the second cases),
$R$ $=$ $R_0$ $=$ 8 $kpc$, the nonsymmetric parts in the term $\Phi^{(2)}$ cause slight oscillations in $R(t)$.

\begin{table}
\small
\begin{center}
\caption{Numerical values of minimum and maximum galactocentric distances
of the Sun under the action of galactic bar, disk and halo.
Motion for the time span of 5 $\times$ 10$^9$ years is considered.}
\begin{tabular}{|c||c|c|c|}
\hline
\multicolumn{1}{|c||}{$v_{c}$} & \multicolumn{1}{|c|}{$v_{y0}$} &
\multicolumn{1}{|c|}{$R_{min}$} &\multicolumn{1}{|c|}{$R_{max}$} \\
\multicolumn{1}{|c||}{[ $km~s^{-1}$ ]} & \multicolumn{1}{|c|}{[ $km~s^{-1}$ ]} &
\multicolumn{1}{|c|}{[ $kpc$ ]} & \multicolumn{1}{|c|}{[ $kpc$ ]} \\
\hline
$240.000$ & $-255.196$ & $7.384$ & $9.193$  \\
$202.176$ & $-220.000$ & $7.774$ & $8.317$  \\
\hline
\end{tabular}
\end{center}
\end{table}

The relative change in $R$ is about $15~\%$ (or even larger) according to
Famaey \& Minchev (2010), Minchev \& Famaey (2010).
Our physical approach yields two different values.
The first case, when $v_{y0}$ $=$ $-255.196$ $km~s^{-1}$, the initial orbital velocity of the Sun,
yields $R_{min} / R_{0}$ $=$ 0.923 and $R_{max} / R_{0}$ $=$ 1.149.
While the value of $R_{max}$ corresponds to the value given by Famaey \& Minchev (2010),
the value of $R_{min}$ is greater than the value presented by other authors.
Moreover, we have to stress that the effect of spiral arms considered by
Minchev \& Famaey (2010), Famaey \& Minchev (2010), Minchev et al. (2010),
Minchev et al. (2011) is not considered in our calculations.
The second case, when $v_{y0}$ $=$ $-220.0$ $km~s^{-1}$, yields
$R_{min} / R_{0}$ $=$ 0.972 and $R_{max} / R_{0}$ $=$ 1.040.
The galactocentric distance of the Sun changes in much smaller interval
than it was presented in the previous papers.
Again, the effect of spiral arms is not considered in our calculations.

Let us consider motion of the Sun during 5 $\times$ $10^{9}$ years.
The first case, characterized by the value $v_{c}$ $=$ $240$ $km~s^{-1}$,
yields that Sun moves within galactocentric distance
$R$ $\in$ $\langle 7.75~kpc, 8.25 ~kpc \rangle$ during 44.3\% of the total integration time,
and, $R$ $\in$ $\langle 7.5~kpc, 8.5 ~kpc \rangle$ during 71.5\% of the total time.
The second case, characterized by $v_{y0}$ $=$ $-220.0$ $km~s^{-1}$,
yields that Sun moves within galactocentric distance
$R$ $\in$ $\langle 7.9~kpc, 8.1 ~kpc \rangle$ during 50.5\% of the total time,
and, $R$ $\in$ $\langle 7.8~kpc, 8.2 ~kpc \rangle$ during 83.2\% of the total time.
The results do not change if we consider the total time span 10 $\times$ $10^{9}$ years
instead of 5 $\times$ $10^{9}$ years.

\begin{table}
\small
\begin{center}
\caption{Numerical values of minimum and maximum galactocentric distances
of the Sun under the action of galactic bar, disk and halo.
Motion for the time span of 5 $\times$ 10$^9$ years is considered,
$v_{y0}$ $=$ $-220.0~km~s^{-1}$.}
\begin{tabular}{|c||c|c|c|}
\hline
\multicolumn{1}{|c||}{$Q_{11} / M$} & \multicolumn{1}{|c|}{$v_{c}$} &
\multicolumn{1}{|c|}{$R_{min}$} &\multicolumn{1}{|c|}{$R_{max}$} \\
\multicolumn{1}{|c||}{[ $pc^{2}$ ]} & \multicolumn{1}{|c|}{[ $km~s^{-1}$ ]} &
\multicolumn{1}{|c|}{[ $kpc$ ]} & \multicolumn{1}{|c|}{[ $kpc$ ]} \\
\hline
$3.4271 \times 10^{6}$ & $202.742$ & $7.752$ & $8.300$  \\
$1.2048 \times 10^{7}$ & $201.669$ & $7.796$ & $8.333$  \\
\hline
\end{tabular}
\end{center}
\end{table}

In reality, there exists some dispersion in the values of $Q_{11}$. Taking into account the
results of Table 1, we can consider the following two extremal values:
$Q_{11} / M$ $=$ 3.4271 $\times$ 10$^{6}$ $pc^{2}$ for the model G1, and,
$Q_{11} / M$ $=$ 1.2048 $\times$ 10$^{7}$ $pc^{2}$ for the model E3.
The results for the radial migration of the Sun are summarized in Table 4.
The Table 4 shows that the values are: $R_{min}$ $\ge$ 7.75 $kpc$ and $R_{max}$ $\le$ 8.33 $kpc$.

\begin{table}
\small
\begin{center}
\caption{Numerical values of minimum and maximum galactocentric distances
of the Sun under the action of galactic bar, disk $+$ halo and spiral arms.
The two-armed spiral structure and $v_{c}$ $=$ 202.176 $km~s^{-1}$,
$v_{y0}$ $=$ $-220.0$ $km~s^{-1}$ and $v_{x0}$ $=$ $0.0$ $km~s^{-1}$ are considered. 
The results hold for the time integration of 5 $\times$ $10^{9}$ years. 
The Sun remains ($T$ / 100) $\times$ 5 $\times$ $10^{9}$ years in the
galactocentric distances $R\in\langle7.8,~8.2\rangle~kpc$.}
\begin{tabular}{|c||c|r|r|c|}
\hline
\multicolumn{1}{|c||}{$\epsilon_{s}$} & \multicolumn{1}{|c|}{$\Omega_{s}$} &
\multicolumn{1}{|c|}{$R_{min}$} &\multicolumn{1}{|c|}{$R_{max}$} &\multicolumn{1}{|c|}{$T$} \\
\multicolumn{1}{|c||}{[ $-$ ]} & \multicolumn{1}{|c|}{[ $v_{y0} / R_{0}$ ]} &
\multicolumn{1}{|c|}{[ $kpc$ ]} & \multicolumn{1}{|c|}{[ $kpc$ ]}& \multicolumn{1}{|c|}{[ $\%$ ]} \\
\hline
\hline
     $$ & $0.7$ & $7.693$ & $8.266$ & $84.9$ \\
     $$ & $0.8$ & $7.759$ & $8.225$ & $93.3$ \\
$0.015$ & $0.9$ & $7.749$ & $8.223$ & $93.1$ \\
     $$ & $1.0$ & $7.758$ & $8.232$ & $94.5$ \\
     $$ & $1.1$ & $7.744$ & $8.224$ & $92.8$ \\
\hline
     $$ & $0.7$ & $7.434$ & $8.445$ & $52.7$ \\
     $$ & $0.8$ & $7.632$ & $8.248$ & $76.5$ \\
$0.030$ & $0.9$ & $7.673$ & $8.197$ & $82.6$ \\
     $$ & $1.0$ & $7.658$ & $8.141$ & $74.9$ \\
     $$ & $1.1$ & $7.696$ & $8.164$ & $85.4$ \\
\hline
     $$ & $0.7$ & $7.109$ & $8.752$ & $39.0$ \\
     $$ & $0.8$ & $7.434$ & $8.362$ & $51.1$ \\
$0.045$ & $0.9$ & $7.519$ & $8.298$ & $60.4$ \\
     $$ & $1.0$ & $7.509$ & $8.225$ & $59.3$ \\
     $$ & $1.1$ & $7.582$ & $8.200$ & $67.5$ \\
\hline
     $$ & $0.7$ & $6.772$ & $8.971$ & $29.3$ \\
     $$ & $0.8$ & $7.201$ & $8.534$ & $37.0$ \\
$0.060$ & $0.9$ & $7.339$ & $8.459$ & $44.2$ \\
     $$ & $1.0$ & $7.333$ & $8.346$ & $44.3$ \\
     $$ & $1.1$ & $7.437$ & $8.268$ & $50.0$ \\
\hline
\end{tabular}
\end{center}
\end{table}

\subsection{Simultaneous effect of the galactic bar and the spiral structure}

If we take into account Eqs. (\ref{corr-spiral arms}), then galactic radial migration of the Sun  is less relevant than Minchev \& Famaey (2010), Famaey \& Minchev (2010), Minchev et al. (2010), Minchev et al. (2011) present. This holds both for the two- and four-armed spiral structures, although
the effect of the two-armed spiral structure is much more relevant than the effect of the
four-armed spiral structure, see Tables 5 and 7.

The results summarized in Tables 6 and 7 show that radial migration of the Sun is practically unaffected by the effect of the four-armed spiral structure with $v_{c}$ $=$ 202.176 $km~s^{-1}$.
The results are practically independent on the spiral structure strength. The found interval
for radial migration of the Sun (7.8 $-$ 8.4) $kpc$ is much shorter than the result
presented by Minchev \& Famaey (2010), Famaey \& Minchev (2010), Minchev et al. (2010), Minchev et al. (2011). We can explain the difference by incorrectness of the equations used by the above mentioned authors. Moreover, it seems to be nonphysical to use the case of resonance $\Omega_{s}$ $=$ $v_{y0} / R_{0}$ as the relevant for the motion of the Sun in the Galaxy. There is no argument why the spiral arms should rotate with the angular velocity equal to the angular velocity of the Sun. The well-known result states that a flat rotation curve exists for the Galaxy. The flat rotation curve yields an orbital speed practically independent on the galactocentric distance and the angular velocity is given by the ratio of the orbital speed and the galactocentric distance. Thus, the angular speed of a galactic object is a function of the galactocentric distance and the fixed value of $\Omega_{s}$, $\Omega_{s}$ $=$ $v_{y0} / R_{0}$, is a pure mathematical toy and it contains no relevant physics.

The last column in Tables 5-7 denotes the relative time $T$ of the existence of the Sun
in the galactocentric distances $R\in\langle7.8,~8.2\rangle~kpc$.
The two-armed spiral structure yields $T$ $<$ 50\% for large values of spiral structure 
strength. The four-armed spiral structure yields $T$ $\in$ (69\%, 79\%).



\begin{table}
\small
\begin{center}
\caption{Numerical values of minimum and maximum galactocentric distances
of the Sun under the action of galactic bar, disk $+$ halo and spiral arms.
The four-armed spiral structure, $v_{c}$ $=$ 202.176 $km~s^{-1}$, 
$v_{y0}$ $=$ $-220.0$ $km~s^{-1}$ and $v_{x0}$ $=$ $0.0$ $km~s^{-1}$ are considered.
The results hold for the time integration of 1 $\times$ $10^{9}$ years.
The Sun remains ($T$ / 100) $\times$ 1 $\times$ $10^{9}$ years in the
galactocentric distances $R\in\langle7.8,~8.2\rangle~kpc$.}
\begin{tabular}{|c||c|c|c|c|}
\hline
\multicolumn{1}{|c||}{$\epsilon_{s}$} & \multicolumn{1}{|c|}{$\Omega_{s}$} &
\multicolumn{1}{|c|}{$R_{min}$} &\multicolumn{1}{|c|}{$R_{max}$} &\multicolumn{1}{|c|}{$T$} \\
\multicolumn{1}{|c||}{[ $-$ ]} & \multicolumn{1}{|c|}{[ $v_{y0} / R_{0}$ ]} &
\multicolumn{1}{|c|}{[ $kpc$ ]} & \multicolumn{1}{|c|}{[ $kpc$ ]}& \multicolumn{1}{|c|}{[ $\%$ ]} \\
\hline
\hline
     $$ & $0.7$ & $7.769$ & $8.297$ & $77.6$  \\
     $$ & $0.8$ & $7.774$ & $8.311$ & $78.6$  \\
$0.007$ & $0.9$ & $7.769$ & $8.308$ & $78.5$  \\
     $$ & $1.0$ & $7.779$ & $8.300$ & $78.2$  \\
     $$ & $1.1$ & $7.772$ & $8.310$ & $78.8$  \\
\hline
     $$ & $0.7$ & $7.764$ & $8.299$ & $72.4$  \\
     $$ & $0.8$ & $7.769$ & $8.328$ & $75.4$  \\
$0.014$ & $0.9$ & $7.766$ & $8.322$ & $74.7$  \\
     $$ & $1.0$ & $7.783$ & $8.309$ & $74.8$  \\
     $$ & $1.1$ & $7.770$ & $8.326$ & $75.8$  \\
\hline
     $$ & $0.7$ & $7.758$ & $8.329$ & $69.9$  \\
     $$ & $0.8$ & $7.765$ & $8.345$ & $73.9$  \\
$0.021$ & $0.9$ & $7.763$ & $8.334$ & $74.2$  \\
     $$ & $1.0$ & $7.781$ & $8.323$ & $73.2$  \\
     $$ & $1.1$ & $7.768$ & $8.339$ & $74.3$  \\
\hline
     $$ & $0.7$ & $7.752$ & $8.360$ & $69.1$  \\
     $$ & $0.8$ & $7.759$ & $8.381$ & $72.7$  \\
$0.028$ & $0.9$ & $7.761$ & $8.357$ & $72.4$  \\
     $$ & $1.0$ & $7.778$ & $8.340$ & $72.9$  \\
     $$ & $1.1$ & $7.763$ & $8.355$ & $73.3$  \\
\hline
\end{tabular}
\end{center}
\end{table}

\begin{table}
\small
\begin{center}
\caption{Numerical values of minimum and maximum galactocentric distances
of the Sun under the action of galactic bar, disk $+$ halo and spiral arms.
The four-armed spiral structure, $v_{c}$ $=$ 202.176 $km~s^{-1}$,
$v_{y0}$ $=$ $-220.0$ $km~s^{-1}$ and $v_{x0}$ $=$ $0.0$ $km~s^{-1}$ are considered.
The results hold for the time integration of 5 $\times$ $10^{9}$ years.
The Sun remains ($T$ / 100) $\times$ 5 $\times$ $10^{9}$ years in the
galactocentric distances $R\in\langle7.8,~8.2\rangle~kpc$.}
\begin{tabular}{|c||c|c|c|c|}
\hline
\multicolumn{1}{|c||}{$\epsilon_{s}$} & \multicolumn{1}{|c|}{$\Omega_{s}$} &
\multicolumn{1}{|c|}{$R_{min}$} &\multicolumn{1}{|c|}{$R_{max}$} & \multicolumn{1}{|c|}{$T$}\\
\multicolumn{1}{|c||}{[ $-$ ]} & \multicolumn{1}{|c|}{[ $v_{y0} / R_{0}$ ]} &
\multicolumn{1}{|c|}{[ $kpc$ ]} & \multicolumn{1}{|c|}{[ $kpc$ ]}& \multicolumn{1}{|c|}{[ $\%$ ]} \\
\hline
\hline
     $$ & $0.7$ & $7.769$ & $8.343$ & $79.9$  \\
     $$ & $0.8$ & $7.769$ & $8.339$ & $80.2$  \\
$0.007$ & $0.9$ & $7.770$ & $8.338$ & $79.9$  \\
     $$ & $1.0$ & $7.776$ & $8.340$ & $79.8$  \\
     $$ & $1.1$ & $7.772$ & $8.333$ & $80.2$  \\
\hline
     $$ & $0.7$ & $7.763$ & $8.372$ & $78.6$  \\
     $$ & $0.8$ & $7.769$ & $8.366$ & $79.6$  \\
$0.014$ & $0.9$ & $7.766$ & $8.361$ & $79.1$  \\
     $$ & $1.0$ & $7.778$ & $8.362$ & $78.3$  \\
     $$ & $1.1$ & $7.767$ & $8.359$ & $78.9$  \\
\hline
     $$ & $0.7$ & $7.758$ & $8.394$ & $77.7$  \\
     $$ & $0.8$ & $7.763$ & $8.396$ & $77.9$  \\
$0.021$ & $0.9$ & $7.764$ & $8.389$ & $77.9$  \\
     $$ & $1.0$ & $7.778$ & $8.385$ & $77.2$  \\
     $$ & $1.1$ & $7.765$ & $8.376$ & $78.5$  \\
\hline
     $$ & $0.7$ & $7.752$ & $8.422$ & $75.7$  \\
     $$ & $0.8$ & $7.756$ & $8.419$ & $75.6$  \\
$0.028$ & $0.9$ & $7.759$ & $8.415$ & $76.1$  \\
     $$ & $1.0$ & $7.771$ & $8.406$ & $76.2$  \\
     $$ & $1.1$ & $7.762$ & $8.395$ & $77.2$  \\
\hline
\end{tabular}
\end{center}
\end{table}

\begin{table}
\small
\begin{center}
\caption{Numerical values of minimum and maximum galactocentric distances
of the Sun under the action of galactic bar, disk $+$ halo and spiral arms.
The four-armed spiral structure and $v_{c}$ $=$ 202.176 $km~s^{-1}$ are considered.
The results hold for the time integration of 5 $\times$ $10^{9}$ years.
The initial $x-$ component of the Sun's velocity is $v_{x}$. The negative sign of
$v_{x}$ corresponds to the orientation toward the center of the Galaxy.
The case when the resonance between the rotation of the Sun and spiral arms is considered:
$\epsilon_{s}$ $=$ 0.028, $\Omega_{s}$ $=$ 1.0 $v_{y0} / R_{0}$,
$v_{y0}$ $=$ 220.0 $km~s^{-1}$ and $R_{0}$ $=$ 8 $kpc$.}
\begin{tabular}{|r||c|c|}
\hline
\multicolumn{1}{|c||}{$v_{x}$}  & \multicolumn{1}{|c|}{$R_{min}$} & \multicolumn{1}{|c|}{$R_{max}$} \\
\multicolumn{1}{|c||}{[$km~s^{-1}$]} & \multicolumn{1}{|c|}{[$kpc$]}   & \multicolumn{1}{|c|}{[$kpc$]} \\
\hline
\hline
$+~10.0$  & $7.558$ & $8.752$  \\
   $0.0$  & $7.771$ & $8.405$  \\
$-~10.0$  & $7.774$ & $8.429$  \\
\hline
\end{tabular}
\end{center}
\end{table}

\begin{table}
\small
\begin{center}
\caption{Numerical values of minimum and maximum galactocentric distances
of the Sun under the action of galactic bar, disk $+$ halo and spiral arms.
The two-armed spiral structure and $v_{c}$ $=$ 202.176 $km~s^{-1}$ are considered.
The results hold for the time integration of 5 $\times$ $10^{9}$ years.
The initial $x-$ component of the Sun's velocity is $v_{x}$. The negative sign of
$v_{x}$ corresponds to the orientation toward the center of the Galaxy.
The case when the resonance between the rotation of the Sun and spiral arms is considered:
$\epsilon_{s}$ $=$ 0.06, $\Omega_{s}$ $=$ 1.0 $v_{y0} / R_{0}$,
$v_{y0}$ $=$ 220.0 $km~s^{-1}$ and $R_{0}$ $=$ 8 $kpc$.}
\begin{tabular}{|r||c|c|}
\hline
\multicolumn{1}{|c||}{$v_{x}$}  & \multicolumn{1}{|c|}{$R_{min}$} & \multicolumn{1}{|c|}{$R_{max}$} \\
\multicolumn{1}{|c||}{[$km~s^{-1}$]} & \multicolumn{1}{|c|}{[$kpc$]}   & \multicolumn{1}{|c|}{[$kpc$]} \\
\hline
\hline
$+~10.0$  & $7.324$ & $8.436$  \\
   $0.0$  & $7.333$ & $8.346$  \\
$-~10.0$  & $7.120$ & $8.667$  \\
\hline
\end{tabular}
\end{center}
\end{table}

\section{Discussion}
The first case treated in the previous section,
$v_{c}$ $=$ $240$ $km~s^{-1}$, corresponds to the value used by, e.g.,
Minchev \& Famaey (2010), Famaey \& Minchev (2010), Minchev et al. (2010),
Minchev et al. (2011). The second approach characterized by
$v_{y0}$ $=$ $-220.0$ $km~s^{-1}$ is the conventional approach, it uses the
conventional speed (e.g., recommended by IAU) of the Sun $220.0$ $km~s^{-1}$.
As a consequence, the radial galactic migration of the Sun is smaller than $\pm$ 0.32 $kpc$
during the interval $(5-10)$ $\times$ $10^{9}$ years.
Thus, the radial migration found by Minchev \& Famaey (2010), Famaey \& Minchev (2010),
Minchev et al. (2010), Minchev et al. (2011) is caused not only by an incorrect
equation of motion, but also by the fact that the effect of galactic bar together with other galactic components is sensitive to the real velocity of the Sun.

Which of the two values of $v_{y0}$ is more realistic?
We can answer this question by calculating the Oort constants.
The first case, corresponding to $v_{c}$ $=$ $240$ $km~s^{-1}$, yields
$A$ $=$ 17.1 $km~s^{-1}~kpc^{-1}$ and $B$ $=$ $-$14.9 $km~s^{-1}~kpc^{-1}$,
if the galactic disk, halo (Eq. \ref{phys-disk+halo}) and the dominant parts of the galactic bar
(Eq. \ref{bar-phi0+kvadr10} -- $\Phi ^{( 0 )} ( \vec{r} )$, $\Phi ^{( 2 )} ( \vec{r} )$) are
considered. The second case, $v_{y0}$ $=$ $-220.0$ $km~s^{-1}$ yields
$A$ $=$ 15.0 $km~s^{-1}~kpc^{-1}$ and $B$ $=$ $-$12.5 $km~s^{-1}~kpc^{-1}$.
The first case is not consistent with the most probably values of the Oort constants,
e.g., Olling \& Merrifield (1998) yield $A$ $=$ (11-15) $km~s^{-1}~kpc^{-1}$,
$B$ $=$ -(12-14) $km~s^{-1}~kpc^{-1}$ (see also Lequeux 2005, p. 10).
Thus, the physical result should correspond to $v_{y0}$ $=$ $-220.0$ $km~s^{-1}$.
Galactic migration of the Sun is less than $\pm$ 0.4 $kpc$ with respect to the current
value 8.0 $kpc$ for the four-armed spiral structure. If we would add a nonzero value of radial velocity component to the velocity of the Sun, then the radial migration might be more relevant, see Table 8. However, the two-armed spiral structure yields much greater radial migration,
up to 1.2 $kpc$ for the largest value of the spiral structure strength.
If we would add a nonzero value of radial velocity component to the velocity of the Sun, then the radial migration might be more relevant, see Table 9.

\section{Conclusion}
The paper presents detailed derivation of the relevant equations of motion
for the Sun under the action of galactic bar and spiral arms. The equations
differ from the equations used by other authors.

Improving physics presented by Dehnen (2000), Minchev \& Famaey (2010),
Famaey \& Minchev (2010), Minchev et al. (2010), Minchev et al. (2011)
and others, we have obtained significantly different radial migration of the
Sun than the previous authors.

Galactic migration of the Sun is less than $\pm$ 0.4 $kpc$, if the four-armed spiral structure 
and the value $v_{y0}$ $=$ $-220.0$ $km~s^{-1}$ are considered. 
The Sun remains (69 $-$ 79)\% of its existence within the galactocentric
distances (7.8 $-$ 8.2) $kpc$ and the result is practically insensitive on the spiral structure
strength. Thus, the radial distance changes only within 5\% from the value of 8 $kpc$.

Galactic radial migration of the Sun is less than $\pm$ (0.3 $-$ 1.2) $kpc$
for the two-armed spiral structure and $v_{y0}$ $=$ $-220.0$ $km~s^{-1}$. 
The Sun remains (29 $-$ 95)\% of its existence
within galactocentric distances (7.8 $-$ 8.2) $kpc$ and the results strongly depend
on the spiral structure strength and the angular speed of the spiral arms.
The radial distance changes within (3.8 $-$ 15.0)\% from the value of 8 $kpc$.


\label{lastpage}

\end{document}